\documentclass[12pt]{article}
\usepackage[export]{adjustbox}
\usepackage{epsfig}
\usepackage[english]{babel}
\usepackage{amsfonts,amsmath,mathrsfs,stmaryrd,amsthm}

\usepackage{rotating}
\usepackage{graphicx}
\usepackage{soul,color}
\usepackage{bm}
\usepackage{float}
\usepackage{tabularx}
\usepackage{booktabs}
\usepackage{epsfig}
\usepackage{natbib}
\usepackage[title]{appendix}
\usepackage{multirow} 
\usepackage{verbatim}
\usepackage{hyperref}
\usepackage{latexsym, alltt}
\usepackage{setspace}
\usepackage{amssymb}
\usepackage{rotating}
\usepackage{tabularx}
\usepackage{caption}
\usepackage[table]{xcolor}
\usepackage{afterpage}
\usepackage{mathtools}
\usepackage{bm}
\usepackage{subcaption}

\usepackage{flafter}
\newcount\Comments  % 0 suppresses notes to selves in text
\usepackage{color}
\definecolor{darkgreen}{rgb}{0,0.5,0}
\definecolor{purple}{rgb}{1,0,1}
\newcommand{\kibitz}[2]{\ifnum\Comments=1\textcolor{#1}{#2}\fi}
\begin{document}
\begin{center}
    {\LARGE Fuzzy Jump Models for Soft and Hard Clustering of Multivariate Time Series Data}\\[1.5em]
    Federico P. Cortese$^{1,2}$, Antonio Pievatolo$^{2}$, Elisa Maria Alessi$^{2}$\\[0.5em]
    {\small $^{1}$Department of Economics, Management, and Quantitative Methods, University of Milan, Milan, Italy}\\
    {\small $^{2}$Institute for Applied Mathematics and Information Technologies ``E. Magenes'', National Research Council of Italy, Milan, Italy}\\[0.5em]
    {\small \texttt{federico.cortese@unimi.it}}
\end{center}

%\date{September  2023}
%\maketitle

\vspace*{-0.5cm}
\begin{abstract}

Statistical jump models have been recently introduced to detect persistent regimes by clustering temporal features and discouraging frequent regime changes.
However, they are limited to hard clustering and thereby do not account for uncertainty in state assignments.
%, which can be problematic in noisy contexts.
%

This work presents an extension of the statistical jump model that incorporates uncertainty estimation in cluster membership. Leveraging the similarities between statistical jump models and the fuzzy $c$-means framework, our \textit{fuzzy jump model} sequentially estimates time-varying state probabilities. Our approach offers high flexibility, as it supports both soft and hard clustering through the tuning of a \textit{fuzziness} parameter, and it naturally accommodates multivariate time series data of mixed types.

Through a simulation study, we evaluate the ability of the proposed model to accurately estimate the true latent‐state distribution, demonstrating that it outperforms competing approaches under high cluster assignment uncertainty.
%
%It also enables clustering of mixed-type data. 
We further demonstrate its utility on two empirical applications: first, by automatically identifying co-orbital regimes in the three‑body problem, a novel application with important implications for understanding asteroid behavior and designing interplanetary mission trajectories; and second, on a financial dataset of five assets representing distinct market sectors (equities, bonds, foreign exchange, cryptocurrencies, and utilities), where the model accurately tracks both bull and bear market phases.

 \noindent \vskip5mm \noindent {

}

%https://www.aeaweb.org/econlit/jelCodes.php?view=jel
%\noindent \vskip5mm \noindent {\sc JEL Classification: }
% D81	Criteria for Decision-Making under Risk and Uncertainty
% C02	Mathematical Methods
% C43	Index Numbers and Aggregation
\end{abstract}

\section{Introduction}

% What is temporal clustering of time series data
Temporal clustering groups multivariate time series that show similar patterns over time. It helps reveal trends, detect anomalies, and improve predictions for future data points \citep{liao:2005}. Clustering methods can be of two types: \textit{hard}, where each observation is assigned to a single group, and \textit{soft}, where each observation can belong to multiple groups with varying degrees of membership.

Among hard clustering approaches for time series data, statistical jump models (JMs) \citep{nystrup:2020} are particularly effective in capturing temporal persistence and in offering an interpretable and explainable model. In the seminal work by \citet{bemporad:2018}, JMs describe complex systems through switches between discrete latent states, each described by a vector of parameters. \citet{nystrup:2020} and \citet{nystrup:2021} extended this framework by integrating 
$k$-means-like clustering and penalizing state transitions to promote persistence, making JMs suitable for high-dimensional applications, particularly in the financial context. \cite{nystrup:2020} show through simulations that JMs are more robust than classical hidden Markov models \citep{bart:farc:penn:13, cortese2024maximum, zucchini:2017}, particularly in challenging settings with limited data, imbalanced regimes, and high persistence.
To clarify, the JMs referenced in this work are distinct from jump-diffusion models, a well-known class of stochastic processes, and should not be confused with them.

JMs have demonstrated high versatility, finding applications across a wide range of domains. In finance, they have been used to analyze cryptocurrency trends \citep{cortese:2023}, model equity market regimes \citep{cortese2024generalized}, for risk management \citep{shu2024downside}, and to inform investment strategies \citep{shu2024dynamic}. Beyond finance, they have been successfully applied to air quality assessment \citep{cortese:2025}, urban thermal comfort monitoring \citep{cortese2024spatio}, and to the study of asteroid dynamics \citep{cortese2025coorbit}.

To address the limitations of hard assignments, \citet{aydinhan2024identifying} recently introduced a \textit{continuous} formulation of JMs, which estimates a probability vector over states at each time point. Estimation is carried out by discretizing the probability simplex with a fixed step size $\delta$, generating a set of candidate probability vectors. The optimal vector at each time point is then selected by minimizing a loss function over this grid using a modification of the \cite{viterbi:1967} algorithm.
While this extension is reliable and efficient, it depends on an exhaustive search over a discretized probability space, whose size grows exponentially with the number of states $K$. Moreover, this method often returns almost hard assignments, even when uncertainty is high.

To overcome these issues, in this work we propose the \textit{fuzzy} JM, a novel method that bridges the gap between hard and soft clustering in the JM framework. By introducing a fuzziness parameter inspired by the fuzzy 
$c$-means algorithm \citep{dunn1973fuzzy,bezdek1981pattern}, our model enables a smooth shift between hard and soft clustering. We estimate state probabilities at each time step through numerical optimization over the probability simplex, following a projected gradient descent procedure \citep{duchi2008efficient}. Following \citet{cortese:2025}, we incorporate the Gower distance \citep{gower1971general} to accommodate both continuous and categorical features. 
%, enhancing robustness and interpretability through state-conditional prototypes. 
This contrasts with earlier JM formulations, which rely on Euclidean distance and are limited to continuous variables. 
%As shown in \citet{cortese:2025}, defining the JM using the $\ell_1$ norm within the Gower distance framework, rather than the standard $\ell_2$ norm, enhances both robustness and clustering accuracy.
%

Through a comprehensive simulation study, spanning a range of sample sizes and number of clusters, we demonstrate that our method accurately recovers true latent‑state probabilities in both hard‑clustering settings (where probabilities approach 0 or 1) and soft‑clustering scenarios marked by high assignment uncertainty. In this latter case, our approach notably outperforms established alternatives such as the $k$‑prototype algorithm of \cite{huang1997clustering} and the continuous JM of \cite{aydinhan2024identifying}.

%\federico{Per EM: Qui sotto aggiungere qualche rif biblio?}
We apply the method to two real datasets. 
%
%One is related to the ephemerides of a real asteroid, which represent the position and velocity of the celestial object over time through variables that describe the geometry of its orbit. 
One is related to the ephemerides of a real asteroid, that represent the state of the celestial object in time through variables that describe the geometry of its orbit.
Asteroids typically transition between distinct orbital regimes, and our approach demonstrates strong capabilities for automatically identifying these transitions. 
%This is particularly relevant in mission planning for interplanetary exploration.
This is important for the understanding of the dynamical evolution of the asteroid population, but also for interplanetary missions.
The other focuses on daily log‑returns of five financial assets. Specifically, we consider three exchange trade funds\footnote{An ETF is a pooled investment instrument that trades on an exchange like a stock and mirrors an index, commodity, or basket of securities.} (ETF) tracking the S\&P500, the U.S. investment‑grade bond market, and the spot price of gold, respectively; additionally, we consider the Bitcoin’s USD price, and the EUR/USD spot rate. These series serve as proxies for the dynamics of equity, fixed‑income, commodity, cryptocurrency, and foreign exchange markets.

We make three main contributions to the current literature.
First, we introduce a novel method for soft clustering of multivariate time series data of mixed type, extending the statistical JM framework to allow for probabilistic state assignments. The proposed approach is computationally efficient, easy to implement, and provides interpretable outputs.
Second, our contribution unifies hard and soft clustering made with JMs within a single general framework. 
%This makes the method particularly suitable for applications where uncertainty in state transitions is intrinsic to the data.
%
%Third, we provide the first structured literature review of statistical JMs, highlighting their connections to traditional clustering methods such as $k$-means and fuzzy $c$-means.
%
Third, we demonstrate the versatility of our method through two applications in distinct domains, namely finance and celestial mechanics. Notably, the latter marks the first application of soft clustering to time series data within this field.
%
%, focusing on a class of orbital dynamics problems. Our results highlight the ability of the proposed method to uncover meaningful temporal structures in complex dynamical systems, demonstrating its potential in both theoretical and applied settings. 

The work is organized as follows. Section \ref{sec:methodology} first presents the fuzzy $c$-means, the statistical JM framework, and the continuous JM, then introduces the proposed model formulation and the algorithm for its estimation. In 
Section \ref{sec:simstud} we demonstrate the strong capability of the proposal to recover the true probability distribution of a mixture model with high persistence through an extensive simulation study.
Section \ref{sec:application} shows the applications of the proposal. Section \ref{sec:conclusions} concludes.

\section{Methodology}
\label{sec:methodology}

In this section, we first introduce the fuzzy $c$-means algorithm, followed by the statistical JM and its recent extension to soft clustering, known as continuous JM. We then discuss the limitations of existing approaches and present our proposed method as a unified framework for both hard and soft clustering of multivariate time series data of mixed type.

\subsection{Fuzzy $c$-means}

%Numerous fuzzy clustering methods have been developed and applied across various domains \citep{durso2019fuzzy}.
%
The \textit{fuzzy} $c$‑means (also known as fuzzy $k$-means) algorithm of \cite{bezdek1981pattern} is the first to introduce a computationally efficient framework for fuzzy clustering. Its justification lies in acknowledging the inherent ambiguity of assigning observations to clusters \citep{maharaj2019time}.

Let $\pmb{Z}\in \mathbb{R}^{T\times P}$, be a  matrix with rows $\pmb{z}_t=\left(z_{t1},\ldots,z_{tP}\right)^\prime \in \mathbb{R}^P$, essentially a collection of $T$ points (possibly representing times) over $P$ features. 
The fuzzy $c$-means clustering method aims at finding $K$ clusters by solving the following optimization problem \citep{durso2015fuzzy,durso2019fuzzy}
\begin{equation}
    \label{eq:fuzzy}
   \underset{\pmb{s}}{\min} \sum_{t=1}^T \sum_{k=1}^K s^m_{tk}d(\pmb{z}_{t},\pmb{\mu}_k)^2 \quad \text{s.t.} \quad \sum_{k=1}^K s_{tk}=1, \,\, s_{tk}\geq 0,
\end{equation}
where $s_{tk}\in [0,1]$ denotes the membership degree of the $t$-th point to the $k$-th cluster, $d(\cdot,\cdot)^2$ is some squared distance  between point $\pmb{z}_{t}$ and centroid $\pmb{\mu}_k$, $k=1,\ldots,K$.

$m>1$ is the parameter controlling the fuzziness of the partition. When $m\xrightarrow{}1$, we obtain standard $k$-means, while $s_{tk}\xrightarrow{} 1/K$ for $m\xrightarrow{}\infty$. 

Considering the Euclidean distance $d(\pmb{z}_{t,},\pmb{\mu}_k)^2=||\pmb{z}_{t}-\pmb{\mu}_k||_2^2$, and solving the constrained optimization method in \eqref{eq:fuzzy} using Lagrangian multipliers, \cite{bezdek1981pattern} show that the solution is given by
\begin{equation}
    \label{eq:sol_fuzz}
    s_{tk}=\frac{||\pmb{z}_{t}-\pmb{\mu}_k||_2^{-\frac{2}{m-1}}}{\sum_{k^\prime=1}^K ||\pmb{z}_{t}-\pmb{\mu}_{k^{\prime}}||_2^{-\frac{2}{m-1}}},
\end{equation}
where the centroids are given by 
$$\pmb{\mu}_{k}=\frac{\sum_{t=1}^T s^m_{tk}\pmb{z}_{t}}{\sum_{t=1}^T s^m_{tk}}.$$
Several extensions of the fuzzy 
$c$‑means to time series data appear in the literature. Among these, \cite{d2009autocorrelation} develop an autocorrelation‑based fuzzy clustering that assigns membership degrees according to auto-correlation function profile of each series. \cite{d2013clustering} propose two GARCH‑based fuzzy models, one using an autoregressive metric, the other a Mahalanobis distance, to capture volatility structures in financial series. \cite{maharaj2011fuzzy} introduce fuzzy clustering in the frequency domain using spectral‐based features that are uncorrelated and therefore well suited to serially dependent time series. \cite{coppi2010fuzzy} extend fuzzy $c$‑means to multivariate spatio‑temporal trajectories by adding a spatial contiguity penalty for improved cluster homogeneity. More recently, \cite{d2021trimmed} present a trimmed fuzzy partitioning-around-medoids approach based on dynamic time warping and outlier trimming, applied to FTSE-MIB constituents.

\subsection{Statistical jump model}

The statistical JM introduced by \citet{bemporad:2018} assumes the existence of a latent state process $\pmb{u}=(u_1,\ldots,u_T)^\prime$, transitioning among $K$ regimes, each characterized by a parameter vector $\{\pmb{\mu}_k\}_{k=1}^K$. Given a collection of time series $\pmb{z}_1, \ldots, \pmb{z}_T$, $\pmb{z}_t \in \mathbb{R}^{P}$, model estimation is performed by minimizing the following loss function
\begin{equation}
    \sum_{t=1}^{T-1} \left[ l(\pmb{z}_t, \pmb{\mu}_{u_t}) + \lambda \mathbb{I}_{\{u_t \neq u_{t-1} \}}\right] + l(\pmb{z}_T, \pmb{\mu}_{u_T}) \,,
\end{equation}
where $l(\cdot,\cdot)$ is a user-defined loss and $\lambda \ge 0$ penalizes transitions between states. As shown in \citet{bemporad:2018}, this framework encompasses a variety of well-known models, including hidden Markov models, depending on the choice of $l$.

\citet{nystrup:2020} apply this model to temporal clustering by 
%transforming the data into standardized $p$-dimensional features, $\widetilde{\pmb{y}}_{t,p} \in \mathbb{R}^p$, and 
adopting a Euclidean quadratic loss. The resulting objective is
\begin{equation}
    \sum_{t=1}^{T-1} \left[  \| \pmb{z}_{t} - \pmb{\mu}_{u_t} \|_2^2 + \lambda \mathbb{I}_{\{u_t \neq u_{t-1} \}} \right] +  \| \pmb{z}_{T} - \pmb{\mu}_{u_T} \|_2^2 \,, \label{eq:JMloss}
\end{equation}
where $\pmb{\mu}_{u_t}$ denotes the mean vector associated with state $u_t$. When $\lambda = 0$, the formulation reduces to standard $K$-means clustering.

Recent extensions of this framework handle mixed‑type data with missing values \citep{cortese:2025} and introduce a spatio‑temporal variant that incorporates a spatial penalty alongside the temporal model to enforce spatial coherence and accommodates irregularly sampled time series \citep{cortese2024spatio}.

\subsection{Continuous jump model}

To overcome the limitations of hard assignments in the discrete jump model, \citet{aydinhan2024identifying} propose a \textit{continuous} extension in which each observation is assigned a probability vector over the $K$ regimes. Let $\pmb{Z} \in \mathbb{R}^{T \times P}$ be the matrix of input features, and let $\pmb{s}_t = (s_{t1}, \ldots, s_{tK})^\prime \in \Delta^{K-1}$ be the probability vector at time $t$, where $\Delta^{K-1}$ is the $K$-simplex.
The model estimates $\pmb{s}_1, \ldots, \pmb{s}_T$ and regime-specific prototypes (in this setting, state-conditional mean vectors) $\pmb{\mu}_1, \ldots, \pmb{\mu}_K \in \mathbb{R}^P$ by solving the following optimization problem
\begin{equation}
\label{eq:cjm}
\underset{\pmb{s}_1,\ldots,\pmb{s}_T,\, \pmb{\mu}_1,\ldots,\pmb{\mu}_K}{\min}
\sum_{t=1}^T \sum_{k=1}^K s_{tk} \, \left\|\pmb{z}_t - \pmb{\mu}_k\right\|_2^2 
+ \lambda \sum_{t=2}^T \left\| \pmb{s}_t - \pmb{s}_{t-1} \right\|_1^2,
\end{equation}
such that $\sum_{k=1}^K s_{tk} = 1$, $s_{tk} \geq 0$.
This loss function consists of a weighted squared distance between each observation and the prototypes, combined with a regularization term that penalizes changes in the state probabilities over time, encouraging persistence.

The optimization is carried out via coordinate descent: for fixed $\pmb{s}_t$, the prototypes $\pmb{\mu}_k$ are updated as weighted averages, while the sequence $\pmb{s}_t$ is estimated via quadratic programming or an approximate dynamic programming scheme.
To estimate $\pmb{s}_t$, the authors essentially discretize the probability simplex by uniformly sampling vectors $c_0, \dots, c_{N-1} \in \Delta^{K-1}$ with grid spacing $\delta > 0$, and collect them into a matrix $C = [c_0 \cdots c_{N-1}] \in \mathbb{R}^{K \times N}$. 
Then they perform a reverse \cite{viterbi:1967} algorithm to reconstruct the sequence of most probable probability vectors. In fact, at each time step $t$, instead of selecting one of $K$ discrete states as in the JM, the method assigns the observation to one of the $N$ candidate probability vectors.

While the continuous JM provides a flexible and reliable modeling framework, it critically depends on the discretization of the simplex. As the grid becomes finer (i.e., as $\delta$ decreases and estimation precision increases), 
the number of candidate vectors grows exponentially with the dimension $K$, making the estimation procedure computationally infeasible in practice.

%\federico{more details on how $s$ is estimated?}

\begin{comment}
    When needed, an additional \textit{mode loss} term can be included to promote sharper (near-hard) assignments:
%
$$
L_\text{mode}(\pmb{s}_{t}) := \log \sum_{j=1}^N e^{-\lambda ||\pmb{s}_t - \pmb{c}_j||_1^2}
$$
%
where $\{\pmb{c}_j\}$ is a grid of candidate points on the simplex.
\end{comment}

% Proposal

\subsection{Fuzzy jump model}

In our framework, the \(P\) features can be continuous or categorical.
Similarly to the fuzzy $c$-means, our \textit{fuzzy} JM produces a sequence of latent state probabilities
%FC sono indeciso fra probabilties e weights :/
\(\pmb{s} = \{\pmb{s}_1, \ldots, \pmb{s}_T\}\), where each \(\pmb{s}_t =(s_{t1},\ldots,s_{tK})^\prime\) is such that $\sum_{k=1}^K s_{tk}=1$, and $s_{tk}\geq 0$, $\forall k$. The second output is a set of state-conditional prototypes \(\pmb{\mu} = \{\pmb{\mu}_1, \ldots, \pmb{\mu}_K\},\, \pmb{\mu}_k=(\mu_{k1},\ldots,\mu_{kP})^\prime\), with prototypes defined as weighted medians for continuous variables and modes for categorical variables. The weighted median is the value that splits the cumulative sum of weights into two equal halves, and the weighted mode is the category with the highest total weight.

We estimate a fuzzy JM
with $K$ states by minimizing the following objective function
\begin{equation}
 \label{eq:fuzzyJM}
f(\pmb{z};\pmb{\mu},\pmb{s}) =    \sum_{t=1}^{T}  \sum_{k=1}^K s_{t,k}^mg(\pmb{z}_{t},\pmb{\mu}_k) 
     +  \sum_{t=2}^{T} \lambda
    %\frac{
    \lVert \pmb{s}_{t-1}-\pmb{s}_{t} \rVert_1^2 \, ,
    %}{\tau_{t+1}-\tau_t} 
\end{equation}
with respect to $\pmb{\mu}=\{\pmb{\mu}_1,\ldots,\pmb{\mu}_K\}$, and $\pmb{s}=\{\pmb{s}_1,\ldots,\pmb{s}_T\}$. 
%
%}

%
%
The function
\(g(\cdot, \cdot)\) 
%is the Gower's distance, a metric for mixed data types: for the detailed formula, we refer to \cite{gower1971general}.
is the \cite{gower1971general} distance, which is a metric used to measure dissimilarity between mixed-type variables.
For two vectors $\pmb{x}_t, \pmb{y}_t \in \mathbb{R}^P$, it is defined as $g(\pmb{x}_t, \pmb{y}_t) = \sum_{p=1}^{P} d_{p}
(\pmb{x}_t,\pmb{y}_t)
%(\pmb{x},\pmb{y})
$
where $d_{p}(\pmb{x}_t,\pmb{y}_t)$ is the contribution of feature $p=1,\ldots,P$ to the distance between $\pmb{x}_t$ and $\pmb{y}_t$, and depends on the type of feature $p$.
%
%\begin{itemize}
   % \item 
    For continuous features 
    \[
    d_{p}(\pmb{x}_t,\pmb{y}_t)
    %(\pmb{x},\pmb{y}) 
    = \frac{|x_{tp} - y_{tp}|}{\sigma_p}\, ,
    \]
    where $\sigma_p$ is a normalizing constant so that $d_p(\cdot,\cdot)\in [0,1]$. 
    We take 
    $\sigma_p=\max(\pmb{x}_{\cdot, p},\pmb{y}_{\cdot,p})-\min(\pmb{x}_{\cdot, p},\pmb{y}_{\cdot,p})$, being the two addend the observed maximum and minimum values of feature $p$ across data points $\pmb{x}_{\cdot, p}=(x_{1p},\ldots,x_{Tp})
    %=(x_{1,p},\ldots, x_{T,p})
    \in \mathbb{R}^T
    $ and $\pmb{y}_{\cdot, p}=(y_{1p},\ldots,y_{Tp})
    %=(y_{1,p},\ldots, y_{T,p})
    \in \mathbb{R}^T
    $.
%
%
    %\item 
    For categorical features
    \[
    d_{p}(\pmb{x}_t,\pmb{y}_t) = 
    \begin{cases} 
    0 & \text{if } x_{tp} = y_{tp}\, , \\
    1 & \text{if } x_{tp} \neq y_{tp}\,.
    \end{cases}
    \]
%\end{itemize}
As shown in \citet{cortese:2025}, defining the JM using the $\ell_1$ norm within the Gower distance framework, rather than the standard $\ell_2$ norm, enhances both robustness and clustering accuracy.

The hyperparameter $\lambda \geq 0$ is a temporal jump penalty that balances data fitting with state sequence stability, where a higher $\lambda$ value makes the model more likely to stay in the same state.

The other hyperparameter \( m \geq 1 \) controls the degree of fuzziness in the clustering process \citep{bezdek1981pattern}. As \( m \to 1 \), the method converges to hard clustering and becomes equivalent to the model proposed by \cite{cortese:2025}.
Conversely, as \( m \to \infty \), the state probabilities approach a uniform distribution of \( 1/K \) for each state.

Equation \eqref{eq:fuzzyJM} corresponds to the objective function of the fuzzy \(c\)-means method proposed by \cite{dunn1973fuzzy} and \cite{bezdek1981pattern} when \(\lambda=0\), which motivates the name we have chosen for our method.

\subsubsection{Estimation}
\label{subsec:estimation}
\cite{nystrup:2021} propose fitting jump models using a coordinate descent algorithm that iteratively alternates between optimizing model parameters for a fixed state sequence and updating the state sequence based on these parameters through the \cite{viterbi:1967} algorithm.  
As already mentioned, \cite{aydinhan2024identifying} estimate state probabilities for a jump model similar to ours using a dynamic programming algorithm, which selects one of \(N\) candidate probability vectors at each time \(t=1,\ldots,T\), with a time complexity of \(O(TN^2)\).  
%While their estimation process is accurate and reliable, it becomes computationally challenging as $N$ increases.

In contrast, we estimate state probabilities for each 
$t$ by solving a constrained optimization problem via \textit{projected gradient descent}, where each iterate of the optimization process is projected onto the probability simplex \citep{duchi2008efficient}. 
%this results in an algorithm with time complexity \(O(TK)\).

%This iterative process is repeated 10 times or concludes earlier if the state sequence stabilizes. 
Global optimality is not guaranteed, so the algorithm is run with 10 initializations, selecting the model with the lowest objective value.

The algorithm can be summarized as follows.
\begin{itemize}
    \item[(a)] Initialize the state probabilities \( \pmb{s} \) (e.g., uniformly) and model parameters \( \pmb{\mu} \) (e.g., using unconditional medians and modes).
    \item[(b)]Iterate the following steps for 
$J$ times or until the objective function changes by less than a given tolerance level.
        %, rather than the highest frequency.
%
\begin{itemize}
    \item[(i)] For each iteration \( j \), sequentially update state probabilities \( \pmb{s}_1^{(j)}, \ldots, \pmb{s}_T^{(j)} \) by numerically solving, for each \( t = 1, \ldots, T \), the following optimization problem

\[
\min_{\pmb{s}_t} \quad \sum_{k=1}^K s_{tk}^m \, g(\pmb{z}_t, \pmb{\mu}_k^{(j)}) + \lambda \cdot \Phi_t(\pmb{s}_t),
\]
subject to
$
\sum_{k=1}^K s_{tk} = 1, s_{tk} \geq 0, \forall k.
$

The regularization term \(\Phi_t(\pmb{s}_t)\) encourages temporal smoothness and is defined differently depending on the position \( t \) in the sequence.
\begin{itemize}
    \item[-] For \( t = 1 \)
\[
\Phi_1(\pmb{s}_1) = \left( \sum_{k=1}^K \left| s_{1k} - s_{2k}^{(j-1)} \right| \right)^2.
\]

\item[-] For \( 1 < t < T \)
\[
\Phi_t(\pmb{s}_t) = \left( \sum_{k=1}^K \left| s_{tk} - s_{t-1\,k}^{(j)} \right| \right)^2 + \left( \sum_{k=1}^K \left| s_{tk} - s_{t+1\,k}^{(j-1)} \right| \right)^2.
\]

\item[-] For \( t = T \)
\[
\Phi_T(\pmb{s}_T) = \left( \sum_{k=1}^K \left| s_{Tk} - s_{T-1\,k}^{(j)} \right| \right)^2.
\]
\end{itemize}
\item[(ii)] Estimate state-conditional prototypes 
%$\pmb{\mu}=\{\pmb{\mu}_1,\ldots,\pmb{\mu}_K\}$
        \begin{equation}
    \label{eq:weightedmean_mode}        \pmb{\mu}^{(j)}=\operatorname*{argmin}_{\pmb{\mu}}\sum_{t=1}^T \sum_{k=1}^K s^{m}_{t,k} g(\pmb{z}_{ t},\pmb{\mu}_{k}).
            \end{equation}
        In the Appendix, we prove that the minimizers are the weighted median and the weighted mode for continuous and categorical variables, respectively.
    \end{itemize}
\end{itemize}

\subsubsection{Hyperparameters selection}
\label{subsec:hyperpar}

\cite{witten2010framework} propose selecting hyperparameters in sparse $k$-means by maximizing the gap statistic, i.e., the difference between the observed between-cluster sum of squares and that from randomly permuted data. Alternatively, hyperparameters can be tuned based on the specific application context. For example, in financial settings, one may use backtesting to select 
$\lambda$ to maximize risk-adjusted returns net of transaction costs \citep{shu2024downside, shu2024Bdynamic, nystrup2019multi, nystrup:2021}. 
Meanwhile, \cite{cortese2023b} and \cite{cortese2024generalized} implement a generalized information criterion for selecting $K$ and $\lambda$ in a statistical JM.
%\federico{Additional comments on this, discuss robustness check as done later}

An additional level of complexity is given here by the presence of the fuzzyness parameter $m$. 
According to \cite{d2015fuzzy}, there seems to
exist no theoretically justifiable manner of selecting it. As a value $m=2$ is often chosen, \cite{chan1992clustering} suggest that this value should be between 1.25 and 1.75.
%\federico{Additional comments on the choice of $m$}
%
%Ozkan and Turksen (2007) indicated upper and lower values for the level of fuzziness inFcM clustering. Based on their analysis, they suggested that the upper boundary value ofthe level of fuzziness should be approximately 2.6 and the lower boundary value approximately 1.4 for FcM clustering in system development practices. For these reasons, theauthors recommended that an analyst should not be concerned about the changes of themembership values outside of these boundaries.
%
%When ground truth labels are available, hyperparameters can be chosen to maximize clustering accuracy. 
%This is the strategy we adopt to select $K, \lambda$ and $m$, using cross-validation (CV) and expert-labeled data as a benchmark.

In this work, we adopt a heuristic approach for hyperparameter selection. The number of states $K$ is fixed a priori guided by theoretical insights from the field. The fuzziness parameter $m$ is chosen to maximize interpretability of the resulting clustering structure; in the case of asteroid data, results remain robust across different values of $m$. For the jump penalty $\lambda$, we select the value at which the estimated fuzzy JM stabilizes, as determined by comparing model outputs for consecutive values of $\lambda$. This procedure enhances robustness with respect to the choice of this hyperparameter. More details on this are given in Section \ref{sec:application}.

\section{Simulation Study}
\label{sec:simstud}

This section aims to evaluate the ability of the proposed fuzzy JM to accurately recover the true time-varying probability distribution over the $K$ clusters.
Specifically,
we consider a generative process in $\mathbb{R}^P$ where each observation $\pmb{y}_t$ arises from a continuous mixture of $K$ multivariate Gaussian components
\begin{equation}
\label{eq:simeq}
    \pmb{y}_t \sim \sum_{k=1}^K \pi_{tk}\, \mathcal{N}_P\bigl(\pmb{\mu}_k, \pmb{\Sigma}_P\bigr),
  \quad \sum_{k=1}^K \pi_{tk} = 1.
\end{equation}
Each centroid $\pmb{\mu}_k$ is defined later according to the value of $K$.
%(e.g., $\pmb{\mu}_1 = (\mu,\ldots,\mu)^\prime$, $\pmb{\mu}_K = (-\mu,\ldots,-\mu)^\prime$, with intermediate values for $2 \leq k \leq K-1$). 
The shared covariance matrix $\pmb{\Sigma}_P$ has unit variances and constant off-diagonal correlation $\rho_{ij} = \rho$ for $i \neq j$.

We obtain states probabilities first considering the latent vector $\pmb{\alpha}_t = (\alpha_{t1}, \ldots, \alpha_{t\,K-1})^\prime$ that follows a vector autoregressive process of order 1,
\begin{equation}
\label{eq:alpha}
    \pmb{\alpha}_t = \boldsymbol{\Phi} \pmb{\alpha}_{t-1} + \pmb{\eta}_t,
  \quad \pmb{\eta}_t \sim \mathcal{N}_{K-1}(\mathbf{0}, \tau^2 I_{K-1}),
\end{equation}
with diagonal autoregressive coefficient matrix $\boldsymbol{\Phi} = \phi I_{K-1}$ and $\phi = 0.99$ reflecting strong temporal persistence in the latent state probabilities. We set $\alpha_{tK} \equiv 0$ for identifiability.

Then, the mixing proportions $\pi_{t,k}$ are generated through a softmax transformation of $K-1$ latent scores $\alpha_{t1},\ldots,\alpha_{t\,K-1}$, 
\begin{equation}
    \label{eq:pi}
    \pi_{tk} = \frac{\exp(\alpha_{tk})}{\sum_{h=1}^{K} \exp(\alpha_{th})}, \quad k=1,\ldots,K.
\end{equation}
This framework induces smooth yet state-dependent transitions between regimes, with the degree of overlap and regime separability modulated by the choice of $\tau$.
Specifically, we consider two scenarios.
\begin{itemize}
  \item \textbf{Soft scenario:} $\tau = 0.2$, leading to more uncertain and smoother state probabilities, typically farther from 0 and 1.
  \item \textbf{Hard scenario:} $\tau = 5$, resulting in sharper state assignments with probabilities pushed toward the boundaries of $[0,1]$.
\end{itemize}

We vary the number of features \(P \in \{5, 10\}\) and the time series length \(T \in \{1000, 2000\}\), as these values are comparable to those observed in the empirical applications of Section \ref{sec:application}.
We simulate 50 independent replicas, always changing the seed, from the generative process in~\eqref{eq:simeq}, estimate a fuzzy JM for each.
We vary $\lambda \in [0,1]$ with step size 0.05, $K \in \{2,3\}$, and $m \in \{1.01, 1.25, 1.5, 1.75, 2\}$. We recall that lower values of $m$ correspond to harder clustering assignments, while higher values yield softer, more uniform probability distributions.

We evaluate performance by comparing the estimated state probabilities with the true ones, as defined in equations \eqref{eq:alpha} and \eqref{eq:pi}, using the average mean squared error (MSE) across all seeds.

We benchmark our method against the continuous JM of~\cite{aydinhan2024identifying}. This comparison justifies the use of continuous features only in this simulation study, since—as previously noted—our method can handle mixed-type data, while the continuous JM is limited to continuous variables. Importantly, clustering performance is not expected to degrade when including categorical features, as demonstrated in the simulation studies of~\cite{cortese:2025}.
In addition, we compare results with the $k$-prototype clustering approach of~\cite{huang1997clustering,huang1998extensions,szepannek2024clustering}, which is a special case of the fuzzy JM when $\lambda=0$ and $m=1$.

We carry out all computations on a 30‑core Intel Xeon Gold 6246R CPU @3.40GHz, where a single fuzzy JM fit requires on average 2.23 minutes. To achieve this performance, the core estimation routines are implemented in \texttt{C++} and then imported in \texttt{R} \citep{R2024} via the \textbf{Rcpp} package \citep{eddelbuettel2011rcpp}. The full source code and instructions are available at the following Github repository \url{https://github.com/FedericoCortese/fuzzyJM.git}.

\subsection{Results}

In the two-regime setting, we define the centroids as $\pmb{\mu}_1 = (1,\ldots,1)^\prime$ and $\pmb{\mu}_2 = (-1,\ldots,-1)^\prime$.
Results in Table~\ref{tab:K2} show that the proposed method outperforms competing models in the soft scenario and achieves similarly low MSE values to the continuous JM in the hard scenario.

The optimal MSE is attained when the number of clusters $K$ is correctly specified, with best results consistently observed at $\lambda \approx 0.40$, $m = 1.01$ in the hard clustering scenario, and $\lambda \approx 0.10$, $m = 1.25$ in the soft scenario.

\begin{table}[h]
\centering
\footnotesize
\caption{
Average mean squared error (MSE) between true and estimated state probability matrices with $K=2$ latent states for the three models: $k$-prototypes ($k$-prot), continuous jump model (cont JM), and fuzzy jump model (fuzzy JM), under the soft and hard scenarios. Monte Carlo standard deviations are in parentheses.
}
\label{tab:K2}
\begin{tabular}{cllllll}
\toprule
& & \multicolumn{2}{c}{Soft Scenario} & \multicolumn{2}{c}{Hard Scenario} & \\
\cmidrule(lr){3-4} \cmidrule(lr){5-6}
$T$ & Method & $P=5$ & $P=10$ & $P=5$ & $P=10$ & \\
\midrule
\multirow{3}{*}{1\,000} 
& $k$-prot & 0.145 (0.016) & 0.151 (0.018) 
             & 0.024 (0.004) & 0.012 (0.004) & \\
& cont JM  & 0.063 (0.014) & 0.063 (0.013) 
             & \textbf{0.009} (0.004) & 0.008 (0.004) & \\
& fuzzy JM & \textbf{0.024} (0.013) & \textbf{0.024} (0.017) 
             & 0.010 (0.004) & \textbf{0.008} (0.004) & \\
\midrule
\multirow{3}{*}{2\,000} 
& $k$-prot & 0.140 (0.011) & 0.144 (0.012) 
             & 0.023 (0.003) & 0.012 (0.002) & \\
& cont JM  & 0.059 (0.010) & 0.057 (0.009) 
             & \textbf{0.008} (0.003) & \textbf{0.008} (0.002) & \\
& fuzzy JM & \textbf{0.022} (0.011) & \textbf{0.020} (0.013) 
             & 0.009 (0.003) & 0.008 (0.003) & \\
\bottomrule
\end{tabular}
\end{table}

In the three-regime setting, we define the centroids as $\pmb{\mu}_1 = (1,\ldots,1)^\prime$, $\pmb{\mu}_2 = (0,\ldots,0)^\prime$, and $\pmb{\mu}_3 = (-1,\ldots,-1)^\prime$.
As in the two-regime case, the lowest MSE is achieved when the number of clusters $K=3$, with optimal performance observed at $\lambda \approx 0.40$ abd $m = 1.01$ in the hard clustering setting, and $\lambda \approx 0.10$ and $m = 1.25$ in the soft clustering setting.
Results in Table~\ref{tab:K3} confirm the strong performance of the proposed method, which outperforms competing models in the soft scenario and achieves MSE values comparable to those of the continuous JM in the hard scenario.

\begin{table}[h]
\centering
\footnotesize
\caption{
Average mean squared error (MSE) between true and estimated state probability matrices with $K=3$ latent states for the three models: $k$-prototypes ($k$-prot), continuous jump model (cont JM), and fuzzy jump model (fuzzy JM), under the soft and hard scenarios. Monte Carlo standard deviations are in parentheses.
}
\label{tab:K3}
\begin{tabular}{cllllll}
\toprule
& & \multicolumn{2}{c}{Soft Scenario} & \multicolumn{2}{c}{Hard Scenario} & \\
\cmidrule(lr){3-4} \cmidrule(lr){5-6}
$T$ & Method & $P=5$ & $P=10$ & $P=5$ & $P=10$ & \\
\midrule
\multirow{3}{*}{1\,000} 
& $k$-prot & 0.122 (0.019) & 0.099 (0.018) 
             & 0.158 (0.048) & 0.122 (0.045) & \\
& cont JM  & 0.076 (0.016) & 0.072 (0.020) 
             & \textbf{0.032} (0.046) & \textbf{0.019} (0.037) & \\
& fuzzy JM & \textbf{0.051} (0.014) & \textbf{0.049} (0.017) 
             & 0.040 (0.044) & 0.032 (0.050) & \\
\midrule
\multirow{3}{*}{2\,000} 
& $k$-prot & 0.112 (0.014) & 0.095 (0.014) 
             & 0.150 (0.043) & 0.113 (0.035) & \\
& cont JM  & 0.073 (0.018) & 0.071 (0.019) 
             & \textbf{0.026} (0.011) & \textbf{0.015} (0.012) & \\
& fuzzy JM & \textbf{0.048} (0.012) & \textbf{0.046} (0.019) 
             & 0.033 (0.013) & 0.024 (0.009) & \\
\bottomrule
\end{tabular}
\end{table}

Taken together, the results in Tables \ref{tab:K2} and \ref{tab:K3} show that the fuzzy JM achieves MSE values 28\% to 50\% lower than those of the continuous JM in the soft scenario, while yielding essentially identical performance in the hard scenario. These findings highlight the greater flexibility of the proposed method and its improved accuracy in settings where cluster assignments are characterized by higher uncertainty.

\begin{figure}[ht]
    \centering
    \begin{subfigure}[b]{\linewidth}
        \centering
        \caption{Soft scenario}
        \includegraphics[width=.8\linewidth] { 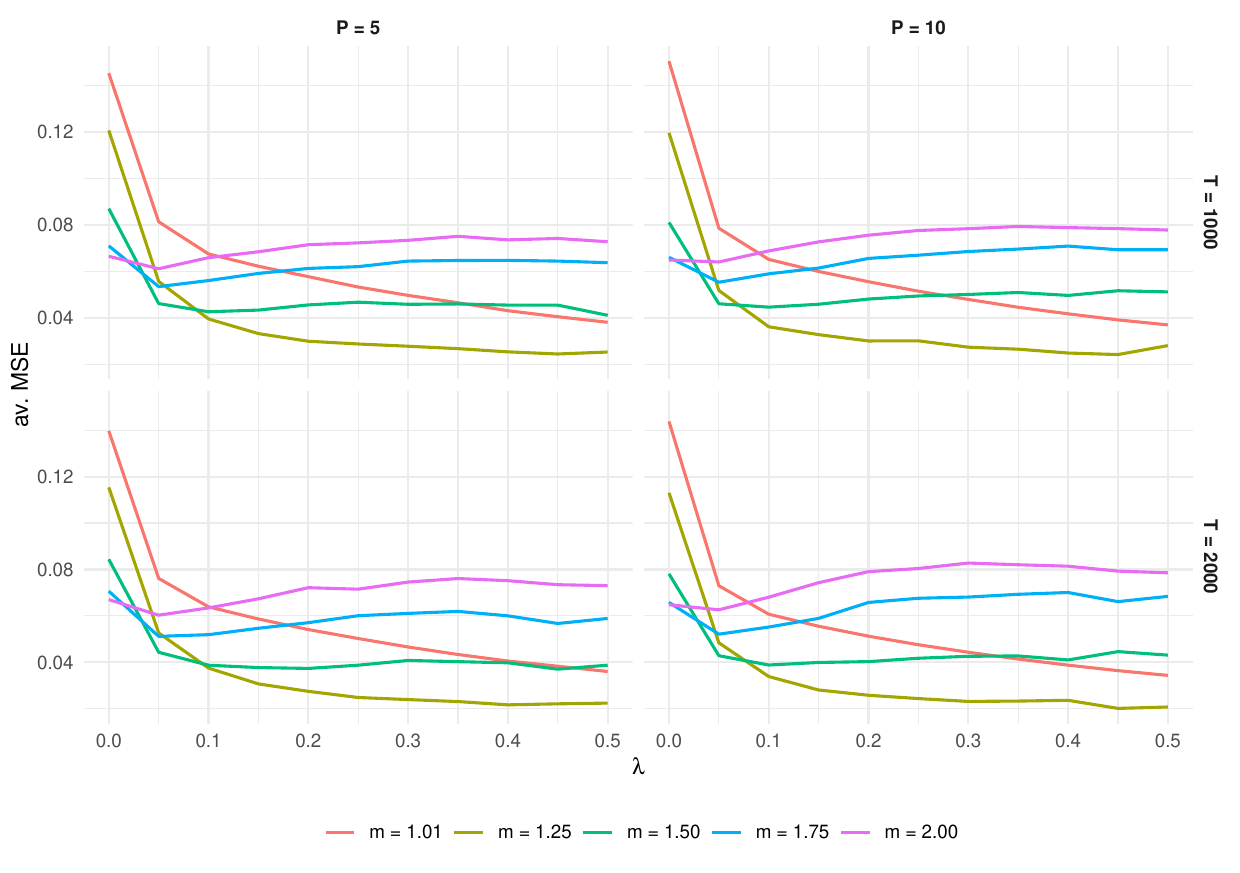}
        %{K2_soft_varLambda.pdf}
        %
        \label{fig:fuzzy_soft}
    \end{subfigure}
    \hfill
    \begin{subfigure}[b]{\linewidth}
        \centering
         \caption{Hard scenario}
        \includegraphics[width=.8\linewidth]
         { 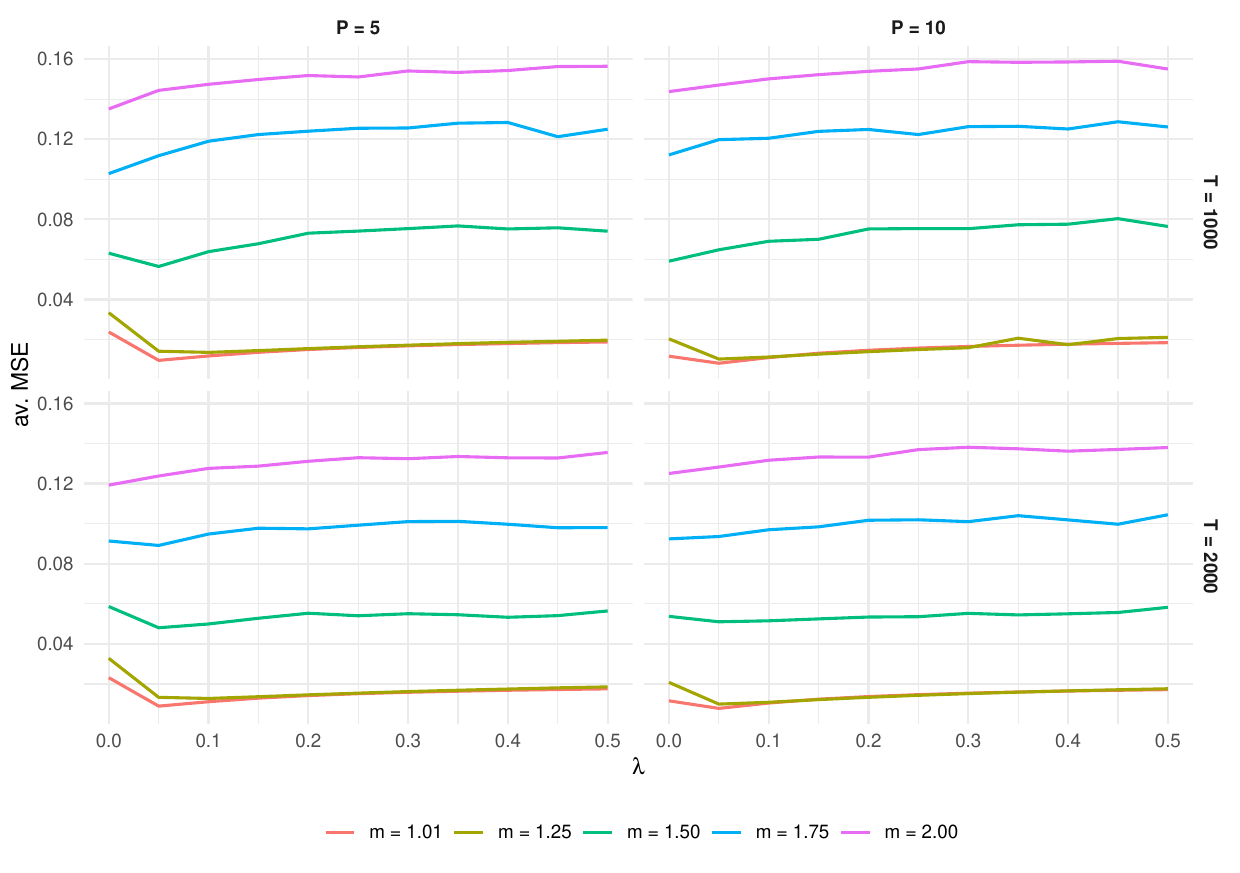}%{K2_hard_varLambda.pdf}
        \label{fig:fuzzy_hard}
    \end{subfigure}
    \caption{Average mean squared error (MSE) between true and estimated state probability matrices with $K=2$ latent states. Panel (a) shows results for the soft scenario, while panel (b) refers to the hard scenario. Each curve represents a different value of $m$ across varying $\lambda$, for different combinations of $T$ and $P$.}
    \label{fig:fuzzy_var_lambdaK2}
\end{figure}

\begin{figure}[ht]
    \centering
    \begin{subfigure}[b]{\linewidth}
        \centering
        \caption{Soft scenario}
        \includegraphics[width=.8\linewidth]{ 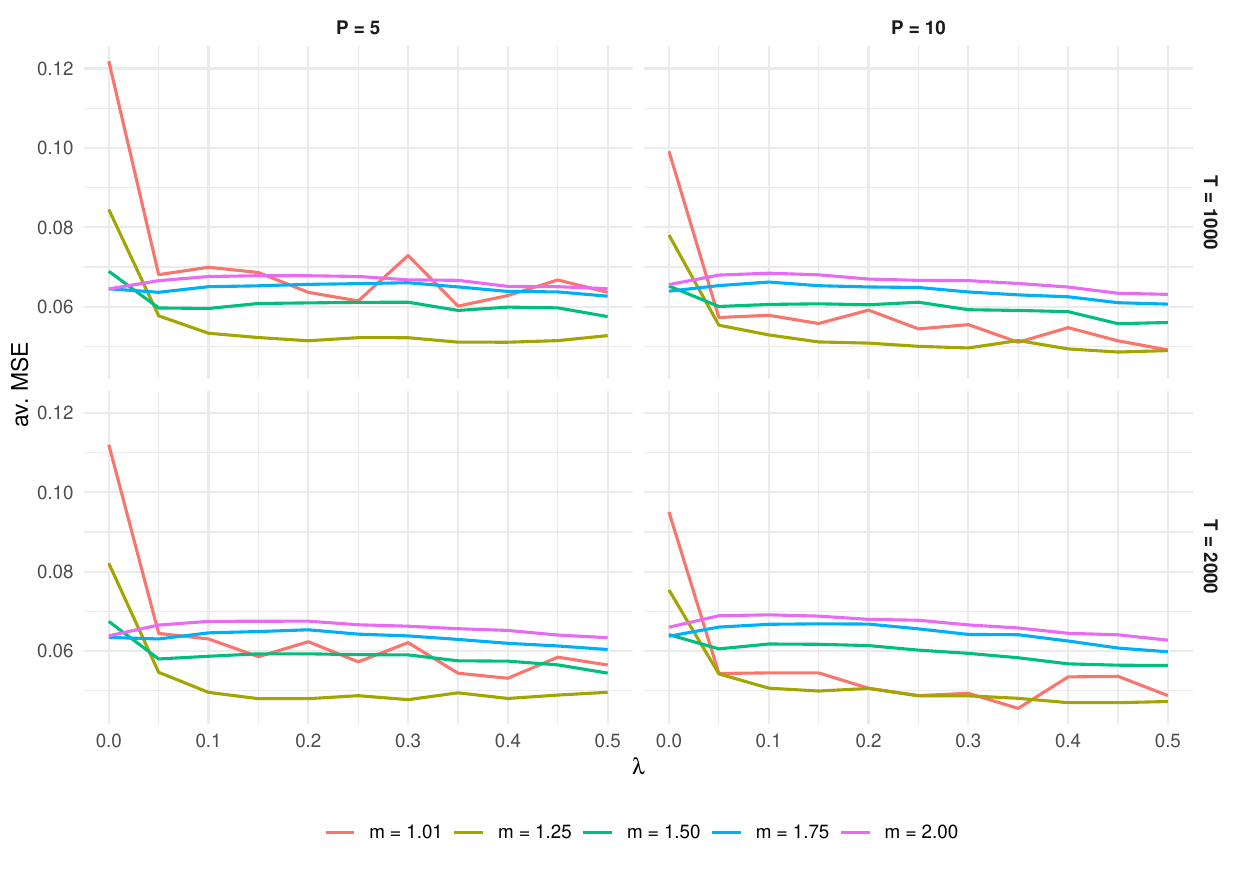}
        \label{fig:fuzzy_soft}
    \end{subfigure}
    \hfill
    \begin{subfigure}[b]{\linewidth}
        \centering
         \caption{Hard scenario}
        \includegraphics[width=.8\linewidth]{ 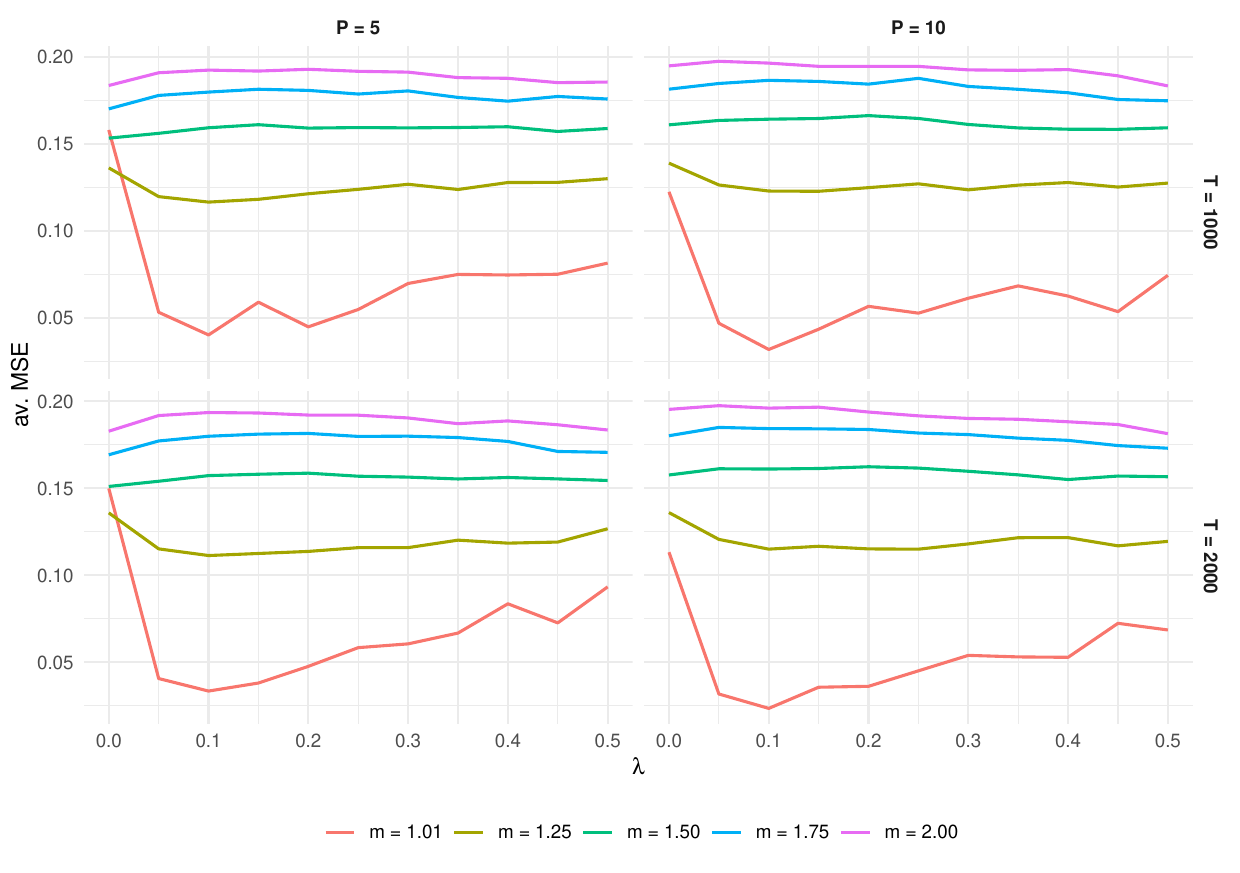}
        \label{fig:fuzzy_hard}
    \end{subfigure}
    \caption{Average mean squared error (MSE) between true and estimated state probability matrices with $K=3$ latent states. Panel (a) shows results for the soft scenario, while panel (b) refers to the hard scenario. Each curve represents a different value of $m$ across varying $\lambda$, for different combinations of $T$ and $P$.}
    \label{fig:fuzzy_var_lambdaK3}
\end{figure}

Figures \ref{fig:fuzzy_var_lambdaK2} and \ref{fig:fuzzy_var_lambdaK3} illustrate how the average MSE varies with $\lambda$ and $m$. The results clearly indicate a preference for the lowest values of $m$ in the hard scenario, and for higher values in the soft scenario. The method also appears robust to the choice of $\lambda$, as performance remains stable across a broad range, with the exception of $\lambda = 0$. As previously noted, this edge case corresponds to the limiting form of the fuzzy JM, which reduces to a standard fuzzy $c$-means.

\section{Applications}
\label{sec:application}

In this section, we present two real‑world applications of the fuzzy JM. 
%First, we analyze data of a real asteroid, with records of its orbital geometry and velocity over time. 
First, we analyze the orbital information of a real asteroid, that describe the time evolution of its geometry.
We demonstrate that the model uncovers the true latent regimes and captures transitions between orbital phases via smoothly varying state probabilities. Second, we apply the method to daily log‑returns of five assets—each representing a different market sector—to illustrate its ability to detect bull and bear phases, with corresponding regime probabilities for each trading day.

\subsection{Asteroid data}

In general, 
%the motion of a celestial body is a conic section, 
a celestial body moves on a conic section, 
defined by five orbital elements describing its size, shape and orientation in a given inertial reference system: namely, \textit{semi-major axis} $a$, \textit{eccentricity} $e$,  \textit{inclination} $i$, \textit{longitude of the ascending node} $\Omega$, \textit{argument of pericenter} $\omega$. 
%A sixth orbital element represents the position of the body along the orbit, for instance the \textit{mean anomaly} $M$.
A sixth orbital element, for instance the mean anomaly $M$, defines the position of
the body along the orbit.

The test case used in this work is the orbital evolution of the asteroid \textit{164207 Cardea}. It is a Near Earth Object, that moves in mean motion resonance with the Earth, meaning that it revolves around the Sun 
%in the same period as the Earth (approximately 1 year). 
in the same period as the Earth does (approximately 1 year).
In celestial mechanics, this specific behavior is known as {\it co-orbital motion} \citep{MoraisMorbidelli2002} and it is characteristic of the so-called ``three-body problem'' \citep{murray1999solar}. There exist different kinds of co-orbital motion depending on the relative phase $\theta$ between the two bodies that orbit around the same central body (i.e., the asteroid and Earth around the Sun).
The relative phase $\theta$ is given by
\begin{equation}\label{eq:theta}
\theta=(M+\Omega + \omega)-(\bar{M}+\bar{\Omega} + \bar{\omega}),
\end{equation}
where $\bar{\Omega}$, $\bar{\omega}$, $\bar{M}$ are the mean anomaly, longitude of the ascending node and argument of pericenter of the heliocentric orbit of the planet.

In particular, we focus on two co-orbital regimes: the horseshoe (HS) regime, where the relative phase \(\theta\) oscillates around \(\pi\) rad, and the quasi-satellite (QS) regime, where \(\theta\) oscillates around 0 radians.
The ephemerides for asteroid 164207, i.e., the time series of its orbital elements and consequently of \(\theta_t\), are obtained from the JPL Horizons system\footnote{\url{https://ssd-api.jpl.nasa.gov/doc/horizons.html}}, which provides accurate orbital evolutions based on real observational data.

As in \cite{cortese2025coorbit}, our goal is to automatically detect transitions between the orbital regimes of the asteroid and provide estimates of the associated uncertainty. To achieve this, we estimate the fuzzy JM on a set of features computed from the data. Specifically, we consider \(P=5\) features derived from \(T=5\,004\) observations. These include the time series of \(\theta_t\) and \(\omega_t\); for each time step \(t\), the closest local minimum \(\text{min}(\theta)_t\) and local maximum \(\text{max}(\theta)_t\) of \(\theta_t\); and the sign of the difference between consecutive values of \(\omega_t\), treated as 
a categorical variable with two levels.

\begin{figure}[ht]
    \centering
    \includegraphics[width=.6\linewidth]{ 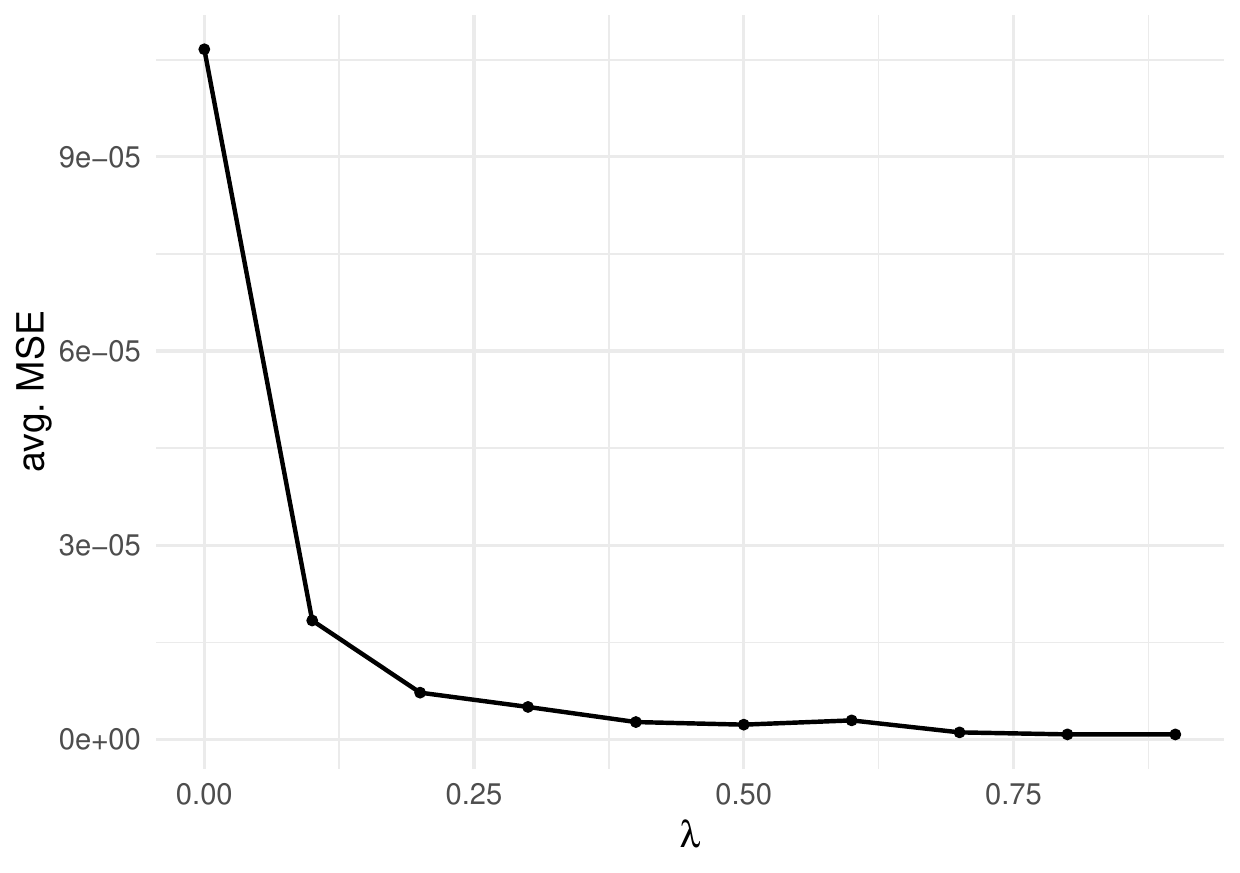}
    \caption{Average mean‐squared error (MSE) between state probability estimates $\hat{\pmb{s}}(\lambda)$ and $\hat{\pmb{s}}(\lambda + 0.1)$, computed for consecutive $\lambda$ values in the interval $[0, 1]$, using the asteroid data.
  %The model is fit with \(K=2\) states and fuzziness parameter \(m=1.1\). 
  %The low, nearly flat curve demonstrates that the inferred state probabilities are largely insensitive to the choice of \(\lambda\).
  }
    \label{fig:MSE_varLambda_asteroids}
\end{figure}

We fix $K=2$ so that the inferred state probabilities can be directly compared with expert‑provided labels at each time step. To select the persistence penalty $\lambda$, we fit a sequence of fuzzy JMs with $K=2$, then compute the average MSE between consecutive soft‑assignment matrices $\hat{ \pmb{s}}(\lambda)$ and $\hat{ \pmb{s}}(\lambda+0.1)$. As shown in Figure \ref{fig:MSE_varLambda_asteroids}, the resulting curve is low and nearly flat for $\lambda\geq 0.50$, indicating robustness to the choice of $\lambda$, so we adopt $\lambda=0.50$.
For what concerns the fuzziness parameter $m$, we choose $m=1.5$ to encourage smoother transitions between states; empirical tests with $m=1.01$ and $m=2$ yield almost the same assignments.

%\subsection{Results}

%Estimation takes 2.23 minutes on a 30 cores Intel (R) Xeon (R) Gold 6246R CPU @ 3.40GHz machine

\begin{figure}[ht]
    \centering
    \includegraphics[width=\linewidth]
    { 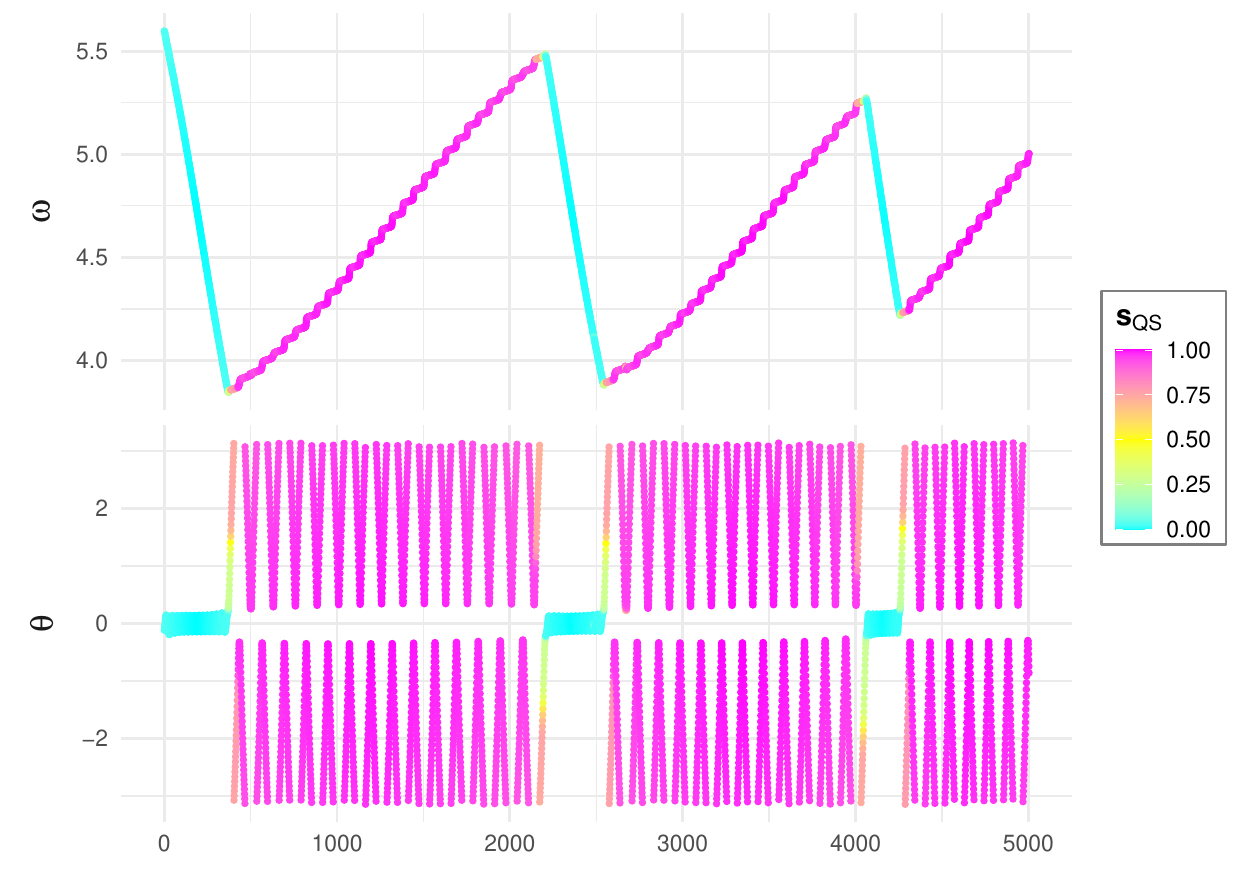}
    \caption{
    Temporal evolution of the argument of pericenter  $\omega_t$ (top) and the relative phase angle $\theta_t$ (bottom), each point colored by the inferred probability of the Quasi‑Satellite (QS) regime $\pmb{s}_{\text{QS}}$. The left y‑axis shows the values of $\omega_t$ and $\theta_t$. Time (in days) is plotted on the x‑axis.
}
    \label{fig:state_class_asteroids}
\end{figure}

Figure \ref{fig:state_class_asteroids} illustrates the time-varying probabilities for the QS regime.  
During transition phases, the probability of switching from HS to QS evolves gradually, showing the ability of the model to anticipate transitions. 
%Transitions from QS to HS occur more abruptly and with a slight advance but are still effectively captured.
%
\label{subsec:results}
Since the data are labeled, we evaluate our classification by comparing the maximum a posteriori (MAP) estimates of the state weights against the manual classification performed by experts. Results show a balanced accuracy of 0.98 and an adjusted Rand index \cite{hubert:1985} of 0.92.  
This result is particularly significant, as manual labeling of dynamical regimes can be extremely time-consuming. In fact, the analyzed series covers approximately $14\,000$ years; manual classification of time series of similar or greater length—common in planetary sciences, where data may span millions of years and involve more complex transitions among multiple co-orbital regimes—is practically infeasible. Thus, such an automated method
%, such as the one proposed here, 
can significantly enhance the efficiency and reliability of co-orbital regime identification.

\subsection{Financial markets data}

The objective here is to track bull and bear phases across major market sectors and to infer how inter‑sector relationships change between these phases. 

We download data from Yahoo Finance through the \textbf{quantmod} \texttt{R} package \citep{ryan2020package} over the period from January 1, 2019, to July 15, 2025. This time span encompasses the COVID‑19 market crash, the 2022 Russian invasion of Ukraine, and the volatility induced by U.S. tariff announcements, essentially three major financial crashes.
Specifically, we consider time series on:
SPY (S\&P500 ETF), which serves as our equity proxy; AGG (iShares Core U.S. Aggregate Bond ETF) capturing the aggregate U.S. fixed‑income universe; GLD (Gold Shares) providing time evolution of gold prices; BTC-USD, the USD price of Bitcoin; and the EUR-USD spot rate, that reflects major foreign exchange dynamics. For each asset we compute two feature sets at daily frequency: (1) log‑returns $\log(P_t) - \log(P_{t-1})$, to quantify directional moves, and (2) a 7‑day rolling standard deviation of those returns, to capture volatility patterns \citep{cortese:2023}. These ten features form the input to our fuzzy JM.

\begin{figure}[ht]
    \centering
\includegraphics[width=.6\linewidth]{ 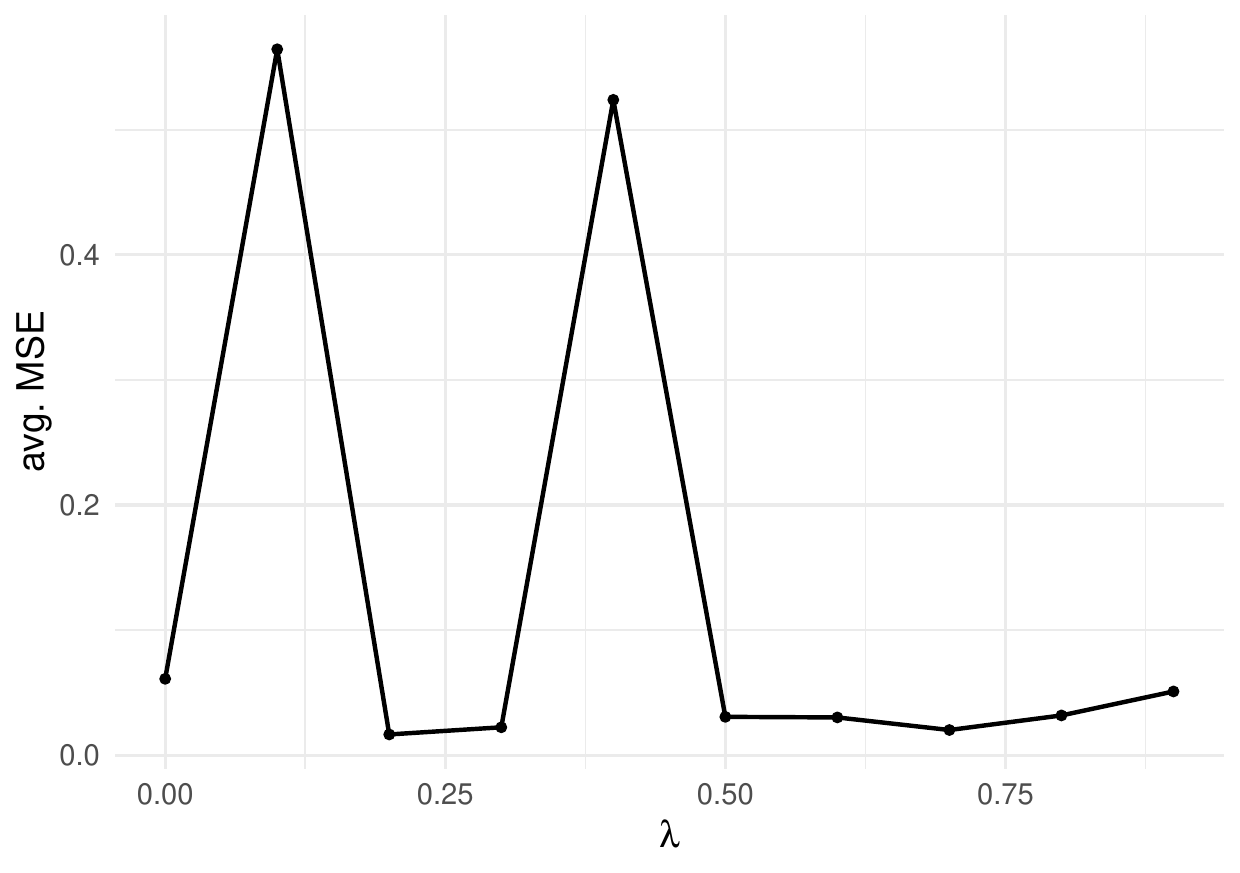}
    \caption{Average mean‐squared error (MSE) between state probability estimates $\hat{\pmb{s}}(\lambda)$ and $\hat{\pmb{s}}(\lambda + 0.1)$, computed for consecutive $\lambda$ values in the interval $[0, 1]$, using the financial markets data.}
    \label{fig:MSE_varLambda_finance}
\end{figure}

We fit the fuzzy JM to our financial feature matrix, holding $K=2$ fixed while varying the persistence parameter $\lambda$ over $\{0,0.1,\dots,1.0\}$. For each $\lambda$, we extract the soft‐assignment matrix $\hat{\pmb{s}}(\lambda)$ and compute the average MSE between $\hat{\pmb{s}}(\lambda)$ and $\hat{\pmb{s}}(\lambda+0.1)$. Figure  \ref{fig:MSE_varLambda_finance} demonstrates that the average MSE remains negligible for $\lambda\geq 0.50$, confirming that the inferred state probabilities are stable across this range. On this basis, we choose $\lambda=0.50$ for all subsequent analysis.

Regarding the choice of $m$, we  consider $m = 1.01$, $m=1.10$ and $m = 1.25$, finding that for $m \geq 1.25$ the state probabilities rapidly converge toward a uniform distribution. This behavior indicates that the effective degree of fuzziness depends not only on $m$ but also on the characteristics of the data. We therefore select $m = 1.1$ to balance a meaningful separation between regimes with clear interpretability of the resulting membership probabilities.

\begin{figure}[ht]
    \centering
    \includegraphics[width=\linewidth]{ 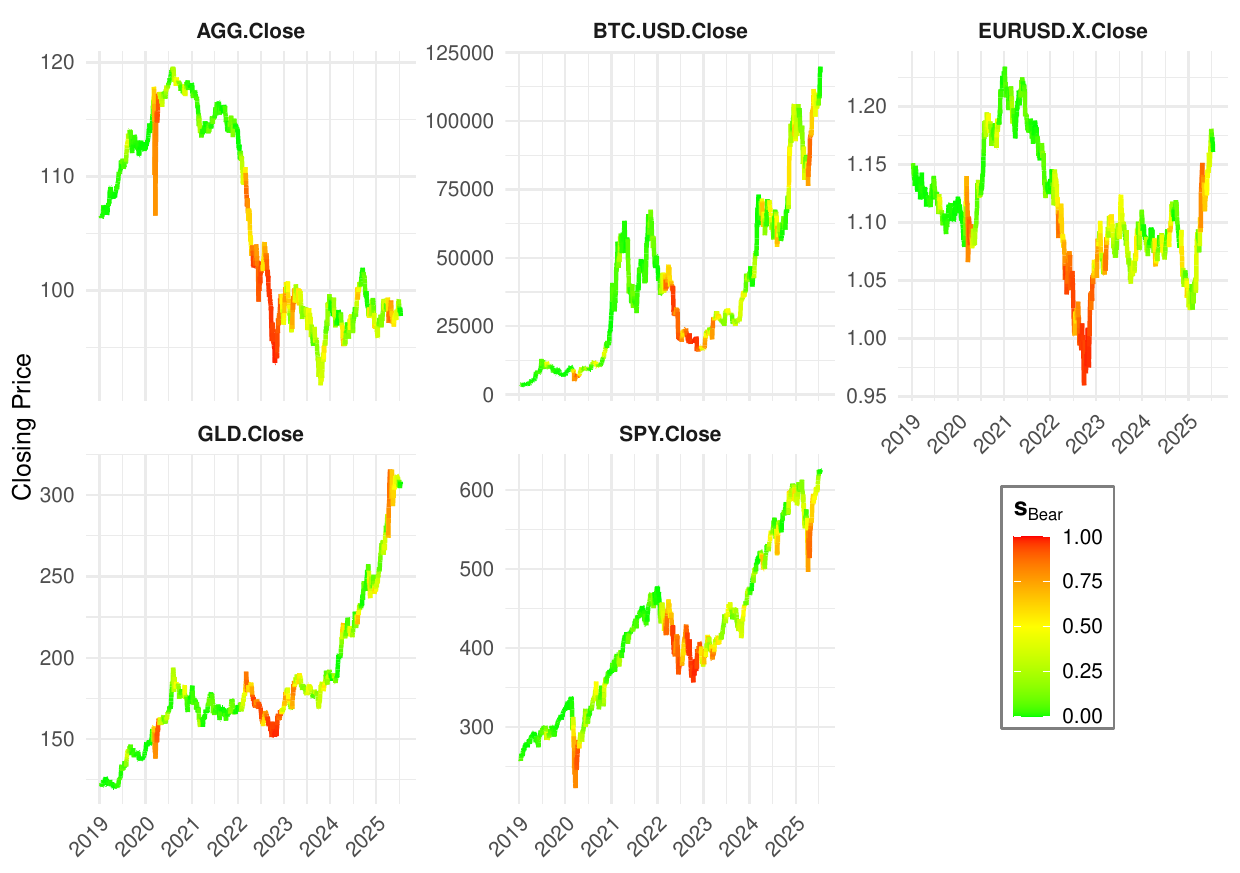}
    \caption{ Daily closing prices of AGG (aggregate bonds), BTC–USD (Bitcoin), EUR/USD spot rate, GLD (gold), and SPY (equities), from January 2019 to July 2025. Each series is colored by the time‑varying probability of the bearish regime, $\pmb{s}_{\text{Bear}}$, as inferred by the two‑state fuzzy jump model: green corresponds to low bear regime probability (bullish conditions), yellow to intermediate, and red to high bear regime probability.  The left y‑axis show the closing prices. Time (in days) is plotted on the x‑axis.}
    \label{fig:finance_appl}
\end{figure}

Figure \ref{fig:finance_appl} shows asset price trends together with the estimated probability of the bear regime over time. Peaks in bear regime probability tend to coincide with major market drops, suggesting that the model effectively captures periods of financial stress.
Additional insights are reported in Table \ref{tab:state_conditional_moments_finance}. State-conditional prototypes are computed via hard assignments, obtained as the maximum a posteriori (MAP) of the estimated probabilities $\pmb{s}_t, \, t = 1,\ldots,T$.
As expected, the bear regime is characterized by higher volatility across all series, particularly for the crypto-asset, and generally lower or near-zero average returns, consistent with periods of market stress or uncertainty. In contrast, the bull regime exhibits lower volatility and higher average returns, especially for the equity index, in line with typical risk-on environments. These patterns reflect well-established stylized facts in financial time series, such as volatility clustering and asymmetric return distributions across market regimes \citep{Hamilton1989, Ang2002}.
\renewcommand{\arraystretch}{1.2}% increase row height
\begin{table}[H]
\centering
\footnotesize
\caption{State‐conditional mean and standard deviation (S.D.) of daily log‐returns (in \%) by regime. 
%Regimes are assigned by the maximum a posteriori of the estimated probabilities \(\pmb{s}_t, \, t=1,\ldots,T\).
}
\label{tab:state_conditional_moments_finance}
\begin{tabular}{clccccc}
\toprule
Regime & Statistic    & AGG    & BTC–USD & EUR/USD & GLD    & SPY    \\
\midrule
\multirow{2}{*}{Bear} & Mean (\%) & –0.016 &  0.008  &  0.023   &  0.068 &  0.008 \\
                      & S.D. (\%) &  0.617 &  4.504  &  0.679   &  1.227 &  1.910 \\
\addlinespace
\multirow{2}{*}{Bull} & Mean (\%) &  0.000 &  0.349  & –0.010   &  0.062 &  0.092 \\
                      & S.D. (\%) &  0.274 &  4.288  &  0.367   &  0.892 &  0.948 \\
\bottomrule
\end{tabular}
\end{table}
Additionally, Table \ref{tab:corr_by_regime} reveals notable differences in cross-sectional correlations between regimes. In the bear phase, correlations are generally stronger, indicating increased comovement during periods of market stress. This behavior aligns with the well-documented phenomenon of contagion effects during crises \citep{Longin2001}. In contrast, the bull regime exhibits weaker overall correlations, suggesting greater potential for diversification.

\renewcommand{\arraystretch}{.65}
\begin{table}[H]
\centering
\footnotesize
\caption{Correlations of daily log‐returns by regime. Values greater than 0.20 are highlighted in bold.}
\label{tab:corr_by_regime}
\begin{tabular}{lccccc@{\quad}ccccc}
\toprule
\multicolumn{6}{c}{Bear regime} 
  & \multicolumn{5}{c}{Bull regime} \\
\cmidrule(r){1-6} \cmidrule(l){7-11}
Asset   & SPY  & AGG  & GLD  & BTC  & EURUSD 
        & SPY  & AGG   & GLD  & BTC    & EURUSD \\
\midrule
SPY     & 1.00 & –    & –    & –    & –     
        & 1.00 & –     & –    & –      & –      \\ 
\addlinespace
AGG     &\textbf{0.27} & 1.00 & –    & –    & –     
        & 0.07 & 1.00  & –    & –      & –      \\ 
\addlinespace
GLD     & \textbf{0.21} & \textbf{0.41} & 1.00 & –    & –     
        & 0.08 & \textbf{0.31}  & 1.00 & –      & –      \\ 
\addlinespace
BTC     & \textbf{0.58} & \textbf{0.34} & \textbf{0.21} & 1.00 & –     
        & 0.17 & –0.00 & 0.11 & 1.00   & –      \\ 
\addlinespace
EURUSD  & 0.14 & 0.19 & \textbf{0.27} & 0.09 & 1.00  
        & 0.10 & 0.19  & \textbf{0.25} & 0.10   & 1.00   \\
\bottomrule
\end{tabular}
\end{table}

\section{Conclusions}
\label{sec:conclusions}

We proposed a novel and interpretable method for soft clustering of mixed-type time series. The approach extended statistical jump models to explicitly account for uncertainty in state classification through an efficient estimation procedure. We demonstrated its effectiveness through an extensive simulation study, where the method achieved higher classification accuracy compared to competing models. Additionally, we showed (i) an application to asteroid ephemerides, where it accurately identified meaningful state transitions and provided promising results in predicting co-orbital motion regimes; and (ii) an application to financial data consisting of five assets from distinct market sectors, where it successfully tracked bull and bear phases and yielded insights into cross-sectional correlations conditional on the latent states.

%%%%%%%%%%%%%%%%%%%
\section*{Appendix}
In this Appendix we prove that weighted median and weighted mode are minimizers of Equation \eqref{eq:weightedmean_mode}. As the Gower distance is the summation of single contributions from each variable, we can split the optimization problem in two parts, one for continuous and one for categorical variables.

\subsection*{Continuous variables}
We aim to minimize the function
\[
f(\mu) = \sum_{i=1}^n w_i |x_i - \mu|.
\]
The derivative of \(f(\mu)\) with respect to \(\mu\) is given by
\[
f'(\mu) = -\sum_{i : x_i \geq \mu} w_i + \sum_{i : x_i < \mu} w_i.
\]
We require \(f'(\mu) = 0\), which implies
\[
\sum_{i : x_i < \mu} w_i = \sum_{i : x_i \geq \mu} w_i.
\]
Now, we relate this condition to the definition of the weighted median.
Define the total weight as
   \[
   W = \sum_{i=1}^n w_i.
   \]
   The equation \(\sum_{i : x_i < \mu} w_i = \sum_{i : x_i \geq \mu} w_i\) implies that \(\mu\) is such that
   \[
   \sum_{i : x_i < \mu} w_i = \frac{W}{2} = \sum_{i : x_i \geq \mu} w_i.
   \]
The weighted median \(\mu^*\) is defined as the smallest value such that
   \[
   \sum_{i : x_i \leq \mu^*} w_i \geq \frac{W}{2},
   \]
   and
   \[
   \sum_{i : x_i < \mu^*} w_i < \frac{W}{2}.
   \]
   So $\mu^*$ is the minimizer.
%
%\item Why This is the Minimizer:
%\begin{itemize}
  %  \item For \(\mu < \mu^*\), the cumulative weight of points less than \(\mu\) is strictly less than \(\frac{W}{2}\). Therefore, \(f'(\mu) > 0\), meaning \(f(\mu)\) is decreasing.
  % \item For \(\mu > \mu^*\), the cumulative weight of points less than \(\mu\) exceeds \(\frac{W}{2}\). Thus, \(f'(\mu) < 0\), meaning \(f(\mu)\) is increasing.
  % \item At \(\mu = \mu^*\), the cumulative weight conditions are satisfied, and the derivative \(f'(\mu)\) transitions from positive to negative. Hence, \(\mu^*\) minimizes \(f(\mu)\).
%\end{itemize}
%

%
%Thus, \(\mu^*\) minimizes the weighted sum of absolute deviations, proving that the weighted median is the solution.

\subsection*{Categorical variables}
Let \(x_i, \, i = 1, \ldots, n\) be categorical variables, with weights \(w_i > 0\). The goal is to minimize the weighted sum of Hamming distances
\[
f(\mu) = \sum_{i=1}^n w_i \delta(x_i, \mu),
\]
where \(\delta(x_i, \mu)\) is defined as
\[
\delta(x_i, \mu) =
\begin{cases}
0 & \text{if } x_i = \mu, \\
1 & \text{if } x_i \neq \mu.
\end{cases}
\]
The term \(w_i \delta(x_i, \mu)\) contributes \(w_i\) if \(x_i \neq \mu\), and 0 otherwise. The total distance is
\[
f(\mu) = \sum_{i : x_i \neq \mu} w_i.
\]
To minimize \(f(\mu)\), we need to minimize the total weight of disagreements. Equivalently, this is achieved by maximizing the total weight of agreements, where
\[
\text{agreements} =f^{C}(\mu)= \sum_{i : x_i = \mu} w_i.
\]
The optimal \(\mu\) is therefore the category that maximizes the total weight
\[
\mu = \underset{c \in \text{Categories}}{\arg \max} \sum_{i : x_i = c} w_i,
\]
which is the definition of weighted mode.

\newpage

%\clearpage
\bibliographystyle{Chicago}
\bibliography{biblio} 

\end{document}